\documentclass{sig-alternate}
\usepackage{balance}
\usepackage{url}
\usepackage{graphicx} 
\usepackage{array} 
\usepackage{algorithm}
\usepackage{algpseudocode}
\newcommand{\nop}[1]{}

\usepackage{color}
\newcommand{\pei}[1]{[Comment from Jian: \textcolor{red}{#1}]}

\renewcommand{\algorithmicrequire}{\textbf{Input:}} 
\renewcommand{\algorithmicensure}{\textbf{Output:}} 
\newtheorem{definition}{Definition}

\newtheorem{lemma}{Lemma}
\DeclareMathOperator*{\Min}{minimize}
\usepackage{times,amsmath,epsfig}

\title{Household Electricity Consumption Data Cleansing}
\numberofauthors{5}
\author{
\alignauthor Guoming Tang\\
			 \affaddr{University of Victoria, Victoria, BC, Canada}\\
       \email{guoming@uvic.ca}
\alignauthor Kui Wu\\
			 \affaddr{University of Victoria, Victoria, BC, Canada}\\
       \email{wkui@uvic.ca}
\alignauthor Jian Pei\\
			 \affaddr{Simon Fraser University, Burnaby, BC, China}\\
       \email{jpei@cs.sfu.ca}
\and
\alignauthor Jiuyang Tang\\
			 \affaddr{National University of Defense Technology, China}\\
       \email{jiuyang\_tang@nudt.edu.cn}  
\alignauthor Jingsheng Lei\\
			 \affaddr{Shanghai University of Electric Power, China}\\
       \email{jshlei@shiep.edu.cn}
}

\begin{document}
\begin{sloppy}
\maketitle
\begin{abstract}
Load curve data in power systems refers to users' electrical energy consumption data periodically collected with meters. It has become one of the most important assets for modern power systems.  Many operational decisions are made based on the information discovered in the data. Load curve data, however, usually suffers from corruptions caused by various factors, such as data transmission errors or malfunctioning meters. To solve the problem, tremendous research efforts have been made on load curve data cleansing. Most existing approaches apply outlier detection methods from the supply side (\textit{i.e.}, electricity service providers), which may only have aggregated load data. In this paper, we propose to seek aid from the demand side (\textit{i.e.}, electricity service users). With the help of readily available knowledge on consumers' appliances, we present a new appliance-driven approach to load curve data cleansing. This approach utilizes data generation rules and a Sequential Local Optimization Algorithm (SLOA) to solve the Corrupted Data Identification Problem (CDIP). We evaluate the performance of SLOA with real-world trace data and synthetic data. The results indicate that, comparing to existing load data cleansing methods, such as B-spline smoothing, our approach has an overall better performance and can effectively identify consecutive corrupted data. Experimental results also demonstrate that our method is robust in various tests. Our method provides a highly feasible and reliable solution to an emerging industry application. 
\end{abstract}


%

\section{Introduction}

Electricity usage data, on the one hand, plays an important role in big data applications, and on the other hand, has been severely under explored.  A recent news article appeared in Forbes~\cite{Forbes} said, ``\emph{But for the most part, utilities have yet to realize the potential of the flood of new data that has begun flowing to them from the power grid, \ldots, And in some cases, they may not welcome it.}'' Yet, existing power grid is facing challenges related to efficiency, reliability, environmental impact, and sustainability. For instance, the low efficiency of current electric grid could lead to $8\%$ of electric energy loss along its transmission lines, and the maximum generation capacity is in use only $5\%$ of the time~\cite{farhangi2010path}. 

The emerging smart grid technology aspires to revolutionize traditional power grid with state-of-the-art information technologies in sensing, control, communications, data mining, and machine learning~\cite{chen2009survey,farhangi2010path}. Worldwide, significant research and development efforts and substantial investment are being committed to the necessary infrastructure to enable intelligent control of power systems, by installing advanced metering systems and establishing data communication networks throughout the grid. Consequently, power networks and data communication networks are envisioned to harmonize together to achieve highly efficient, flexible, and reliable power systems. 

Among the various types of data transmitted over the smart grid, load curve data, which refers to the electric energy consumption periodically recorded by meters at points of interest across the power grid, has become the critical assets for utility companies to make right decisions on energy generation, billing, and smart grid operations. Load curve data, which ``is beginning to give us a view of what the customer is actually experiencing, something that we've never ever seen before''~\cite{Forbes}, is precious user behavior data, and is an important type of big data.

Load curve data collected and reported from smart meters at end-users' premises is especially important for both energy supply and energy demand sides. On the demand side, it has direct impact on customers' energy bills and their trust on the still nascent smart grid technology. On the energy supply side, inaccurate load data may lead to large profit losses and wrong business decisions. In $2012$, $126.8$ million residential customers in the US used over $1,374$ billion kWh, which counts to over $33\%$ of the total electric energy in the US~\cite{EIA}. The importance of this huge amount of energy and its financial implication cannot be over emphasized. 

Nevertheless, it is \textit{\textbf{unavoidable}} that load curves contain corrupted data and missing data, caused by various factors, such as malfunctioning meters, data packet losses in wireless networks, unexpected interruption or shutdown in electricity use, and unscheduled maintenance~\cite{chen2010automated}. Due to the huge volume of load curve data, it is hard for utilities to manually identify corrupted load curve data. Unfortunately, problems caused by corrupted data are usually realized only after it is too late, such as after a customer receiving a suspicious yet hard to rebut high energy bill. 

As a concrete example, according to the news reports~\cite{bchydro1,bchydro2}, some customers in the province of British Columbia, Canada, were baffled by energy bills that are more than double what they were charged before the smart meter installation. While the problem could be identified by common sense and certain agreement might be reached by good faith negotiations~\cite{bchydro1,bchydro2}, fixing the questionable bill is another head-scratching and embarrassing issue to the utility. As a response to customer complaints, the utility normally took remedy actions, such as replacing the smart meters or taking back the smart meters for lab testing~\cite{bchydro2}. Such a remedy, however, can hardly be effective. According to CBC News~\cite{CBC}, ``Government estimates indicate there are about 60,000 smart meter holdouts (in the province).'' Overall, the users and the utility company have the well-aligned interest and should work together to tackle this critical problem plaguing the electric power industry.

\nop{Technically, load curve data is usually transmitted over multi-hop wireless communication networks in a wide area. The reliability of wireless networks is subject to many environmental factors, such as sporadic radio interference or even different weather conditions, all hard to duplicate in lab testing.  Overall, the users and utility companies have the well-aligned interest and should work together to tackle this critical problem plaguing the electric power industry.   }

Techniques of load data cleansing have been proposed to deal with load data corruption problem recently~\cite{chen2010automated}. Most existing load data cleansing methods are designed for the supply side (\textit{i.e.}, electricity service providers), to help the utility companies find the corrupted data and protect their profits. From the supply side, the collected load data is usually aggregated data, \textit{i.e.}, the energy consumption of a billing unit such as a house or a commercial building. When performing data cleansing on the supply side, due to the difficulty of obtaining extra knowledge behind the aggregated load data, most existing approaches apply outlier detection methods, \textit{i.e.}, the data that deviates remarkably from the regular pattern is identified as corrupted data. Various assumptions about the data generation mechanism are required for outlier detection, but due to limited information, those assumptions are usually based on empirical knowledge or statistic features of the data. Such outlier detection methods are oblivious of appliances' various energy consumption models and may not be accurate or fair to customers. We call these methods \textit{appliance-oblivious}.  Such methods suffer from a few important deficiencies.  

For example, the regression-based outlier detection methods find statistical patterns of load data and claim the data significantly deviating from the patterns as corrupted data. Nevertheless, such resulted outliers are not necessarily corrupted data. In addition, without the knowledge of appliances' energy consumption models, some ``hidden" corrupted data is hard to detect.  To be specific, the energy consumption of a group of appliances in a house or a building is a stochastic process. The stochastic feature makes it hard to establish a fixed pattern. Turning on/off any high-power appliance may lead to a steep change in load curve. Using appliance-oblivious data cleansing methods, the data generated under such a condition is likely to be captured as outliers.

As another example, appliance-oblivious methods cannot deal with ``hidden'' corrupted data. Fig.~\ref{fig:motiveExp} shows an example of three appliances, $A_1$, $A_2$, and $A_3$, which have power ranges of $[2,4],[10,12]$ and $[30,32]$, respectively. The load data within some ranges such as $(4,10),(16,30)$, and $(36,40)$ cannot be generated by any combination of the three appliances. Nevertheless, such data may not be identified by existing outlier detection as corrupted data.

\begin{figure}[t]
\begin{center}
\includegraphics[width=0.45\textwidth]{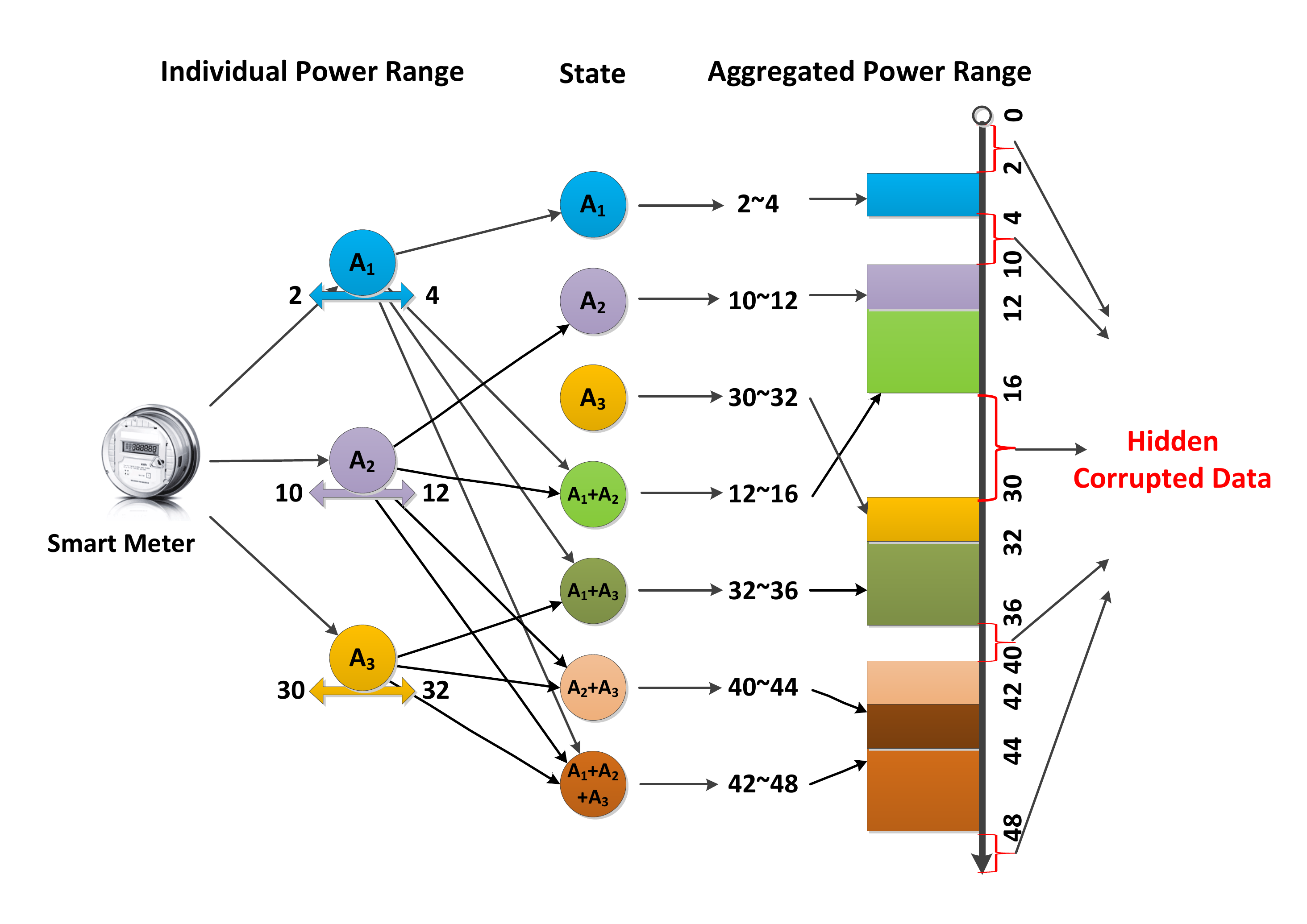}
\caption{An example showing hidden corrupted data generated with three appliances}\label{fig:motiveExp}
\end{center}
\end{figure}

\nop{Recently, with the emergence of smart appliances and fine-grained in-house energy monitoring systems, customers now have the capability to monitor their own energy usage more closely and more accurately~\cite{Stragier10,Dong13}. As such, }

Recently, with the emergence of fine-grained in-house energy monitoring systems, customers now have the capability to monitor their own energy usage more closely and more accurately~\cite{Stragier10}. As such, users may possess more knowledge behind the data, \textit{e.g.}, the decomposition of total energy consumption according to main appliances. Even if in-house energy monitoring system is not available, users should know the rated power of appliances', which are easily accessible from the appliances' manual, technical specification or public websites, such as~\cite{EnergyStar2013}. This knowledge presents new opportunities to perform load data cleansing on the customer side directly or on the supply side with auxiliary information from customers. This new angle of tackling the corrupted load data problem can greatly improve the quality of load data. 


In this paper, we tackle the practical problem of corrupted load curve data identification in the industry context by developing an appliance-driven approach.  Specifically, we make the following contributions:
\begin{itemize}
	\item First, we define a new criterion in identification of corrupted load data. The criterion is aware of domain knowledge, including the power range of appliances and the physical laws behind valid load data. 
	\item Second, we formally formulate the \textit{Corrupted Data Identification Problem} (CDIP) and establish an optimization model to solve the problem. Furthermore, we introduce a new concept, called \emph{virtual appliance}, in the objective function to help record corrupted data. Our empirical study in a proof-of-concept electricity usage test environment shows that a solution to the optimization problem is capable of precisely identifying the corrupted data, even without obtaining the exact on/off states of appliances. This nice feature indicates that our method is both effective and robust.
	\item We develop a \emph{sequential local optimization algorithm} (SLOA) to approach CDIP efficiently. SLOA focuses on solving CDIP in a smaller time window, and considers the correlation between consecutive small windows. While we show that the original CDIP problem is NP-complete, our SLOA method offers an efficient heuristic solution and can achieve a very high detection precision. As an extra benefit, by applying the sequential optimization algorithm, we can easily handle consecutive corrupted data.
\end{itemize}
The rest of the paper is organized as follows. In Section~\ref{sec:relatedwork}, we review the related work. In Section~\ref{sec:problemDefine}, we formally formulate the corrupted data identification problem (CDIP). To solve CDIP, we develop an optimization model in Section~\ref{sec:systemModel}. As CDIP is proven to be NP-complete, we develop SLOA to find an approximate solution in Section~\ref{sec:solutions}. We evaluate the performance of our method with real-world trace data and synthetic data in Section~\ref{sec:evaluation} and Section~\ref{sec:evaluation2}, respectively, and test the robustness of SLOA in Section~\ref{sec:robust}. The paper is concluded in Section~\ref{sec:conclusion}.

\section{Related Work}\label{sec:relatedwork}

\nop{Because of the importance of load data to both demand side and supply side, load data cleansing has caught more and more attention in recent smart grid research. So far,}Most related literature treats the corrupted data the same as outliers in load pattern and focuses on outlier detection. A broad spectrum of techniques for outlier detection in load data have been developed, which include regression-based time series analysis, univariate statistical methods, and data mining techniques.  We review them briefly here.

\subsection{Regression-Based Time Series Analysis}
Regression-based time series analysis is the most widely used approach for outlier detection in load data~\cite{abraham1989outlier,chen2010automated,ljung1993outlier, mateos2012robust}. Mateos and Giannakis~\cite{mateos2012robust} developed a nonparametric regression method that approximates the regression function via $\ell_{0}$-norm regularization.  Chen~\textit{et al.}~\cite{chen2010automated} proposed a nonparametric regression method based on B-spline and kernel smoothing to identify corrupted data. Abraham and Chuang~\cite{abraham1989outlier} analyzed residual patterns from some regression models of time series and used the patterns to construct outlier indicators. They also proposed a four-step procedure for modeling time series in the presence of outliers. Greta \textit{et al.}~\cite{ljung1993outlier} considered the estimation and detection of outliers in time series generated by a Gaussian auto-regression moving average (ARMA) process, and showed that the estimation of additive outliers is related to the estimation of missing observations. ARMA is also utilized in~\cite{fox1972outliers, abraham1988score, abraham1989outlier, schmid1986multiple} as a fundamental model to identify outliers.

\subsection{Univariate Statistical Methods}
Univariate statistical methods are another type of techniques for outlier detection. Univariate statistical methods deal with outliers in load data by processing load data as one-dimensional real values~\cite{ferguson1961rejection, david1979robust, gather1989testing, davies1993identification}. Most univariate methods for outlier detection assume an underlying \textit{a prior} distribution of data. The outlier detection problem is then translated to finding those observations that lie in the so-called outlier region of the assumed distribution, which is defined by a confidence coefficient value~\cite{davies1993identification}. Since the statistical methods are susceptible to the number of exemplars, a simple but effective method named Boxplot or IQR is proposed in~\cite{tukey1977exploratory} to deal with small-sized exemplars. 

\subsection{Data Mining Techniques}
In addition to the above methods, data mining techniques are also applied to identify outliers, such as k-nearest neighbor~\cite{ramaswamy2000efficient, knox1998algorithms}, k-means\nop{~\cite{allan1998topic}}, k-medoids~\cite{bolton2001unsupervised}, and DBSCAN~\cite{kriegel2005density}.  As a type of clustering methods, they group data with similar features, and identify data items that do not strongly belong to any cluster or far from other clusters as outliers. Recently, Aggarwal~\cite{Agga13} provided a thorough survey on outlier detection.

Nevertheless, all the above methods do not consider the special physical laws behind the load data. Regression-based methods assume that the data follows a certain pattern, which can be modeled by a function governed by a set of parameters; univariate methods assume that the data is sampled from a certain known distribution; clustering methods assume that the data is well structured as clusters and the corrupted data deviates significantly from the normal structure. Obviously, the underlying assumptions in the existing methods are quite general and do not capture the specific features of load data well. 
Our paper fills the gap and differs from the existing literature by offering a completely new angle to address the load data corruption problem.

\section{The Corrupted Data Identification Problem}\label{sec:problemDefine}

In this section, we present a formal problem definition.  Before that, we first describe an energy consumption model and discuss the generation rules of load data.

\subsection{Energy Consumption Model}
Load data is time series data that records users' energy consumption. It is collected by smart meters \textit{periodically} at a certain sampling frequency. Without loss of generality, we assume that the time is slotted, with each timeslot equal to the sampling interval time. In the rest of the paper, we thus use the terms ``\textit{time}'', ``\textit{timeslot}'' and ``\textit{sampling interval}'' interchangeably. 

We denote the load data from timeslot $t = 1$ to timeslot $t = n$ in a column vector as
 \begin{equation}\label{loadData}
Y \equiv [y_1,y_2,\cdots,y_n]^T,
\end{equation}
where each value $y_i$ in the vector represents the aggregated energy consumption of all appliances in a property, say a house at timeslot $i$. The energy consumption at a time instant depends on the appliances' on-off states and their individual power level.  

We assume that a house includes $m$ appliances in total, and the power of the $k$-th appliance is $p_k$ (\textit{watts}). At any time instant, if we record the power level of each individual appliance, we can define an $m$ dimensional column \emph{power vector} to capture energy consumption of the house: 
\begin{equation}\label{powerVector}
P \equiv [p_1,p_2,\cdots,p_m]^T.
\end{equation}

Note that the power level of an appliance normally does not remain at a fixed value but changes in a certain range. For this reason, we define two $m$ dimensional column vectors, denoted as $P_l$ and $P_u$, respectively:
\begin{align}
P_l = [l_1,l_2,\cdots,l_m]^T \\
P_u = [u_1,u_2,\cdots,u_m]^T,
\end{align}
where $l_i$ and $u_i$ represent the lower and upper bounds of the power level of the $i$-th appliance, respectively. A power vector $P$ is called \emph{valid} if, for each value $p_i$ in $P$, $l_i \le p_i \le u_i$. 

At any instant, the state of an appliance may be either \textit{on} or \textit{off}. We use an $n\times m$ $0$-$1$ \emph{state matrix}, $S=[S_{ij}]_{n\times m}$, to record the states of the $m$ appliances from time $t=1$ to $t = n$, where $S_{i,k}=1$ indicates that the $k$-th appliance is on at time $i$, and $0$ otherwise. In addition, we call the $i$-th row of $S$ a \textit{state vector} at time $i$, denoted by:
\begin{equation}
S_i \equiv [S_{i,1},S_{i,2},\cdots,S_{i,m}].
\end{equation}

\subsection{Generation Rules of Load Data}

We have the following observations. First, a valid load data element $y_i$ (in \textit{watt-hours}) should be equal to the inner product of the state vector and the power vector at $t=i$, multiplied by the sampling interval time. This is a basic physical law for load curve data generation. Second, since the sampling interval is typically small (\textit{e.g.}, $10$ seconds\footnote{Residential smart meters can support sampling rate as high as $1$ sample per second~\cite{Weiss}.}), we assume that the probability that an appliance has more than one on-off switch events during a timeslot is negligible. In addition, the total number of on-off state switches of all appliances during a timeslot should be small. This feature is called the \textit{temporal sparsity} of on-off state switching events. Intuitively, this feature means that in normal operation it is unlikely that the household turns on-off many appliances in a short time. Based on the above observations, we can define the generation rules of load data. 
\begin{definition}[Generation Rules] Assume that the initial state of appliances is $S_{0}$. We claim that each valid load data, $y_{i}$, must satisfy the following rules:

\begin{equation}\label{Eqt:generRule}
 \left\{ \begin{array}{ll} 
S_{i}\cdot P_{l}/f \leq y_{i} \leq S_{i}\cdot P_{u}/f \\
\lVert S_{i} - S_{i-1} \rVert_1 \leq \delta,
\end{array} \right.
\end{equation}
\noindent where $f$ is data sampling frequency, $1\leq i \leq n$, and $\delta$ is the upper bound on the total number of on-off state switches for $m$ appliances during a sampling interval.    
\end{definition}

Note that the energy consumption value (watt-hours) is calculated with power value (watt) multiplied by time $1/f$ (hour). To keep our discussion simple, we assume that a valid initial state vector $S_0$ is given at this moment. We will relax this assumption later and show that the impact of an inaccurate initial state vector quickly becomes negligible as long as the system runs for just a little while (Section~\ref{sec:robust}).  

\subsection{Problem Definition}

Based on the above generation rules, \emph{corrupted data} is the values that break any of the rules. 

\begin{definition}[CDIP] The \textbf{corrupted data identification problem (CDIP)} is, given load data $Y=\{ y_1,y_2,\cdots,y_n\}$, power bound vectors $P_{l}, P_{u}$, and a sampling frequency $f$, find corrupted data items that violate any of the generation rules, \textit{i.e.}, $C \equiv \{y_{i}:y_{i} \text{ violates } (Equation~\ref{Eqt:generRule}),\text{ for } 1 \leq i \leq n\}$.
\end{definition}

\section{An Important Step Towards Solving CDIP}\label{sec:systemModel}


To solve CDIP, a na\"{\i}ve idea is to find all the solutions satisfying the constraints in (Equation~\ref{Eqt:generRule}), by brute-force, exhaustive search for all possible appliance states. This method is very time-consuming.  Even for a small data set it is very costly to find the answer. Since the generation rules can be considered as constraints in an optimization problem, we will show how the problem can be transformed to an optimization problem, which sheds light on a fast solution to an approximate problem. 

\begin{definition}[Virtual Appliance] Besides the real appliances, we introduce a virtual appliance into the system. Its associated power is called \textit{virtual power}, and we record the values of virtual power from time $t=1$ to $t=n$ in a \textit{virtual power vector}
\begin{equation}
V \equiv [v_1,v_2,\cdots,v_n]^{T},
\end{equation}
\noindent where $v_i\in (-\infty, +\infty)$ denotes the virtual power at time $t=i$.
\end{definition}
\nop{Since it is hard to solve $\ell_{0}$-norm in practice, $\ell_{1}$-norm is usually taken as a good alternative. It is proven that $\ell_{1}$-norm not only gives a near-optimal sparsest solution for $\ell_{0}$-norm, but also has some good properties that can be applied to develop efficient solutions~\cite{donoho2006most}. In addition, as a convex expression, $\ell_{1}$-norm can be processed by many off-the-shelf optimization tools such as \emph{CVX}~\cite{cvx}.\nop{ \emph{Gurobi}and \emph{Mosek}}. For the above reasons, we re-formulate the problem from Equation~\ref{minimization1} as:}
\nop{Note that $v_i$'s essentially are slack variables in the optimization problem. They come into play whenever the generation rules cannot be satisfied with real appliances.} 

By introducing the virtual appliance, we can develop the following optimization model to solve CDIP:
\begin{equation}\label{minimization3}
\begin{aligned}
& \Min_{S_i, v_i}
& & \left\| V \right\|_{1} \\
& \text{subject to}
& & \left( S_i\cdot P_{l} + v_i \right)/f \leq y_i \leq\left(S_i\cdot P_{u} + v_i \right)/f\\
&&& \lVert S_{i} - S_{i-1} \rVert_{1} \leq \delta \\
&&& S_{i,j} \in \{0,1\} \\
&&& 1\leq i \leq n \\
&&& 1\leq j \leq m 
\end{aligned}
\end{equation}

To understand the rationale behind the formulation of Equation~(\ref{minimization3}), it is worthwhile to point out that $v_{i}\in V$ will come into play whenever $S_{i}$ cannot satisfy the generation rules, \textit{i.e.}, the virtual appliance is ``turned on'' when the load data $y_{i}$ is corrupted. Thus, $v_{i}$ essentially makes a record to the corrupted data. After obtaining the final solution to Equation~(\ref{minimization3}), the $v_i$ variables with non-zero values indicate the corrupted data, \textit{i.e.}, 
\begin{equation}
C = \{y_i: v_i \neq 0 \text{ for } 1 \leq i \leq n\}.
\end{equation}
We try to minimize $\ell_1$-norm, because it is proven that for most large under-determined systems of linear equations the minimal $\ell_1$-norm solution is also the sparsest solution (i.e., resulting in the minimal number of non-zero values of $v_i$)~\cite{donoho2006most}. In addition, a larger $v_i$ value means that a corrupted $y_{i}$ is farther away from a valid range. In this sense,  the value of $v_i$ can be also regarded as the \textit{corrupted degree} of $y_i$. 

The problem can be proven NP-complete (refer to Appendix A). By investigating the special structure of the problem, however, we can develop an effective heuristic algorithm introduced in the next section.

\section{Sequential Local Optimization Algorithm}\label{sec:solutions}

In this section, we propose a \emph{Sequential Local Optimization Algorithm} (SLOA) and develop a quantitative strategy to estimate the minimum local window size.

\subsection{SLOA}
The temporal sparsity of corrupted load data suggests that we can perform optimization in a smaller, local time window. By considering the correlation between consecutive timeslots, we design a Sequential Local Optimization Algorithm (SLOA). Without loss of generality, we take a load data from time $t = 1$ to $t = n$ as an example to show the major steps of SLOA.
\begin{enumerate}
\renewcommand{\theenumi}{Step \arabic{enumi}}
\item \label{varient} Consider a small time window with size of $w, 1 \leq w < n$, which starts from time $k$ to time $k+w-1$. Given the state vector at time $k-1$, \textit{i.e.}, $S_{k-1}$, we consider the following optimization problem:
\begin{equation}\label{minimization4}
\begin{aligned}
& \Min_{S_i, v_i}
& & \sum_{i=k}^{k+w-1}| v_i | \\
& \text{subject to}
& & \left(S_{i}\cdot P_{l} + v_{i}\right)/f \leq y_i \leq \left(S_{i}\cdot P_{u} + v_{i}\right)/f \\
&&& \lVert S_{i} - S_{i-1} \rVert_{1} \leq \delta \\
&&& S_{i,j} \in \{0,1\} \\
&&& k  \leq i \leq k + w -1 \\
&&& 1\leq j \leq m
\end{aligned}
\end{equation}
By setting $w \ll n$, we can significantly reduce the search space. Actually, we can show that the computational complexity to solve the above problem is $O(M^{w})$, where $M={{m}\choose{0}}+{{m}\choose{1}}+\cdots+{{m}\choose{\delta}}$ (refer to Appendix B). Since $m$ is the total number of appliances and $\delta \ll m$, the problem can be solved quickly using tools such as \emph{CVX} $2.0$ with a \emph{Gurobi} engine~\cite{cvx}.

\item \label{continous} For the $k$-th time window that starts from time $k$, we use the following strategy to handle consecutive corrupted data: if the data point at time $k$ is identified to be corrupted, \textit{i.e.}, $v_k \neq 0$, recover the current state vector from the previous one, \textit{i.e.}, set $S_k = S_{k-1}$.

\item Repeat~\ref{varient}) and~\ref{continous}) from $k = 1$ to $k = n$, and solve problems in form of Equation~(\ref{minimization4}) sequentially. After $n$ iterations, we can get a sequential solution $v_1,v_2,\cdots,v_n$. Thus, the corrupted data set is $C = \{y_{i}:v_{i} \neq 0, 1 \leq i \leq n\}$, in which $v_i$ is the corrupted degree of load data $y_i$.
\end{enumerate}

Algorithm~\ref{alg:1} shows the pseudo code of SLOA.
\begin{algorithm}[t]
\caption{Sequential Local Optimization Algorithm}\label{alg_seqentialOpt}\label{alg:1}
\begin{algorithmic}[1]
\Require 
Load data $\{y_{1},y_{2},\cdots,y_{n}\}$, power bounds $P_l,P_u$, initial state $S_0$, sampling frequency $f$, local time window size $w$.
\Ensure 
Corrupted data set $C$, corrupted degree $v_{i}, 1 \leq i \leq n$

\State $v_0 = 0$
\State $C = \varnothing$
\For {$k = 1:n$}
	\State Solve Problem (Equation~\ref{minimization4}), and obtain $v_i$ and $S_i$ where $k \leq i \leq k+w-1$
	\If {$v_k \neq 0$}
			\State $C = C \cup \{y_k\}$
			\State $S_k = S_{k-1}$
	\EndIf
\EndFor
\State \textbf{return} $C, \{v_1,v_2,\cdots,v_n\}$
\end{algorithmic}
\end{algorithm}

\subsection{Estimation of Minimum Local Window Size}
Clearly, one key question in SLOA is to determine a suitable size of the local window. In principle, we want the size to be as small as possible to speed up the calculation, but a size too small may result in a poor solution largely deviating from the global optimal one. For example, in the extreme case of $w=1$, SLOA becomes a simple greedy search algorithm, where the final solution may not be good. On the other hand, if $w=n$, the problem becomes the same as Equation~(\ref{minimization3}), which is hard to solve. \textit{What is the minimum local window size that leads to a nearly global optimal solution?}

Since it is hard to obtain a strict proof that the local optimal solutions together lead to the global optimal solution, we use the following heuristics to estimate the minimum local window size: within the local window, it should be possible that one state vector can be transited to any other state vectors in the vector space. In other words, within the local window, we should cover all possible state vectors in the search. This heuristics sheds light on finding the minimum local window size, as formulated below.       

\begin{definition}[Overlap Index]
Consider $m$ appliances denoted by a set $\{R_{1},R_{2},\cdots,R_{m}\}$, where $R_{i} \equiv [l_i, u_i] \subset \Re$ and represents the $i$-th appliance's power range. The overlap index of $m$ appliances is defined as:

\begin{equation}
O \equiv \frac{\sum^{m}_{i=1}\int^{p_{max}}_{p_{min}}{I\left(R_{i}\cap \{x\}\right)}dx}{\int_{P_{min}}^{P_{max}}I( (\cup_{i=1}^m R_i)\cap\{x\})dx}
\end{equation}

\noindent where $p_{max}$ and $p_{min}$ stand for the maximum and minimum power of all appliances, respectively, and $I(x)$ is an indicator function
\begin{equation}
I(x) = \left\{ \begin{array}{ll} 
1 \text{ , } x\neq \varnothing\\
0 \text{ , } x = \varnothing
\end{array} \right.
\end{equation}
\end{definition}

Note that the denominator $\int_{P_{min}}^{P_{max}}I( (\cup_{i=1}^m R_i)\cap\{x\})dx$ includes all valid power values, \textit{i.e.}, the ones that can be covered by at least one appliance's power range. We can see that the overlap index represents the number of appliances whose power range covers a valid power value, averaged over the whole power range of all appliances. In particular, $O = 1$ indicates that no pair of appliances have overlapped power, and $O=m$ means that all appliances have the same power range. Intuitively, when $O$ is large, there is a good chance of finding multiple local optimal solutions to Equation~(\ref{minimization4}), since there are multiple equivalent choices to turn on/off appliances in each iteration. 

With the heuristics in estimating the minimum local window size, we have the following result. 
 \begin{lemma}\label{lemma:1}
Given the overlap index $O$ of $m$ appliances and the upper bound $\delta$ on the total number of on-off state switches in a timeslot, in order to get the nearly global optimal solution to Equation~(\ref{minimization3}) via Equation~(\ref{minimization4}), the minimum local window size $w = max\{\left\lceil \frac{m}{\delta\cdot O}\right\rceil , 1\}$.
\end{lemma}
\proof We prove the following condition holds: within the local window, we can cover all possible state vectors in the search.

First, the value of $w$ relates to upper bound $\delta (\leq m)$ on the total number of on-off switches in one timeslot. It is obvious that from time $t=i$ to $t=i+1$, the state vector $S_i$ can only change to another state vector $S_{i+1}$, with $\lVert S_{i+1} - S_{i} \rVert_{1} \leq \delta$. If $\delta = m$, then within one step, a state vector is allowed to change to any other state vector. On the other hand, if $\delta=1$, within one step, a state vector can only change one value in the vector. In other words, from one state vector, it requires at least $\left\lceil \frac{m}{\delta}\right\rceil$ timeslots to reach any other state vector in the state vector space, \textit{i.e.}, $w \geq \left\lceil \frac{m}{\delta}\right\rceil$.        

Second, the overlap index $O$ can reduce the value of $w$. Based on the meaning of $O$, $m$ appliances with overlap index $O$ are equivalent to $\left[m/O\right]$ appliances without overlapped power. Replace $m$ with $m/O$, we can get $w \geq \left\lceil \frac{m}{\delta\cdot{O}}\right\rceil$. Considering $w\geq 1$, we conclude:
\begin{equation}\label{equ:windowSize}
w \geq max\{\left\lceil \frac{m}{\delta\cdot{O}}\right\rceil , 1 \}.
\end{equation}
Since we want $w$ to be as small as possible, $w=max\{\left\lceil \frac{m}{\delta\cdot{O}}\right\rceil , 1 \}$.
$\Box$

\nop{With our setting of local window size, the more ambiguous of the appliances is, the smaller window size is, and the lower computational complexity problem (Equation~\ref{minimization4}) will be. Fortunately, we have found that, in practice of most situations, $w_{min}$ can be really small, sometimes even as smaller as one, which reduces algorithm (Equation~\ref{alg_seqentialOpt}) to a simple greedy method.}

Please note that, although the minimum local window size obtained above is an estimation, it works effectively in our experiments with real-world data as well as with synthetic data.  

\subsection{Algorithm Analysis}

Given $n$ load values, $m$ appliances, and the upper bound $\delta (\leq m)$ on the total number of on-off state switches in a timeslot, the computational complexity of the original problem (Equation~\ref{minimization3}) is $O(M^{n})$, where $M={{m}\choose{0}}+{{m}\choose{1}}+\cdots+{{m}\choose{\delta}}$. Using SLOA, solving the optimization problem (Equation~\ref{minimization4}) for $n$ times results in the time complexity of $O(n \cdot M^{w})$, where $w \in Z^{+}$ and $w \ll n$ (refer to Appendix B). Please note that the appliance number $m$ is a constant value and $w$ is also a small constant.

Obviously, the larger the value of $w$, the higher the computational complexity. Fortunately, the overlap index of appliances in a house or a building is usually high, as observed in our real-world experiment testbed. This fact allows us to select a small local window size following Lemma~\ref{lemma:1}. Therefore, in the application scenarios, SLOA can approach the NP-complete problem heuristically and efficiently.  We will show that this algorithm indeed can provide a good solution with abundant experimental results in the following sections.

\section{Experimental Evaluation on Real Data}\label{sec:evaluation}

We evaluate our method with real-world trace data from a real-world energy monitoring platform. We monitored the appliances' energy consumption of a typical laboratory and a lounge room on the fifth floor of the Engineering/Computer Science Building at the University of Victoria (UVic) for two months. The real-time power of laptops, desktops and some household appliances was recorded. Each appliance's power level was measured every $10$ seconds and the measurement results were transmitted with ZigBee radio to a server that stores the data. The monitored appliances and their regular power\footnote{An appliance's regular power is an approximate range around the rated power where this appliance works.} are shown in Fig.~\ref{fig_appDeployment}.

\begin{figure}[t]
\begin{center}
\includegraphics[width=2.5in,height=2.0in]{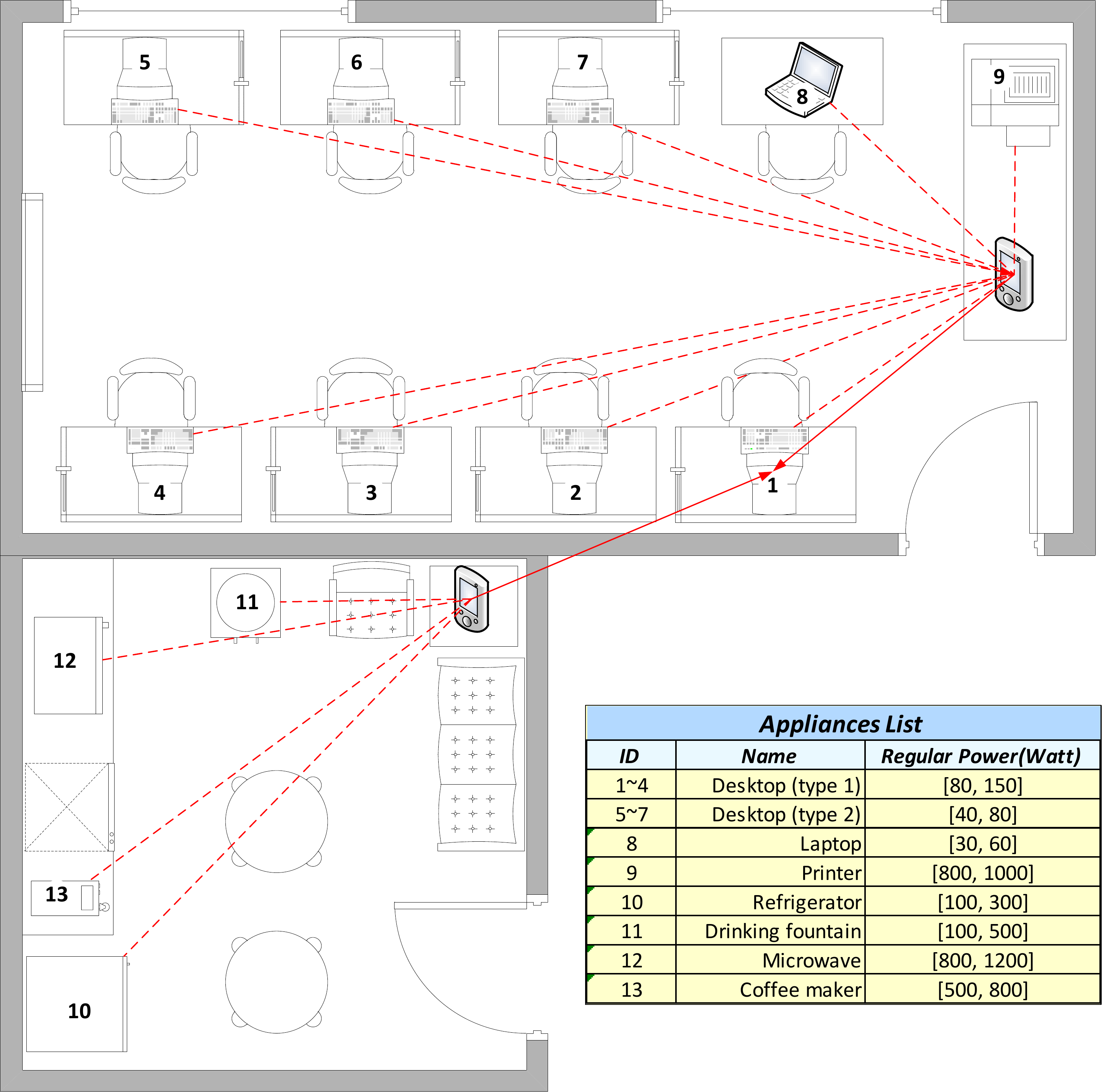}
\caption{Energy monitoring platform and appliances' power ranges}\label{fig_appDeployment}
\end{center}
\end{figure}


We test the data day by day over the two-month period. Fig.~\ref{fig_realdataExp} demonstrates one-week and one-day load data collected by our platform. For clear illustration, we only show one-day data as an example. Note that even in a lab setting like ours, there indeed exists some apparent corrupted data, indicated by the dashed red dots in Fig.~\ref{fig_realdataExp}\footnote{The corrupted data mainly comes from some incorrect power values from the laptop that occasionally reports impossible values such as hundreds of Watts.}. 

\begin{figure}[t]
\begin{center}
\includegraphics[width=3.4in,height=1.6in]{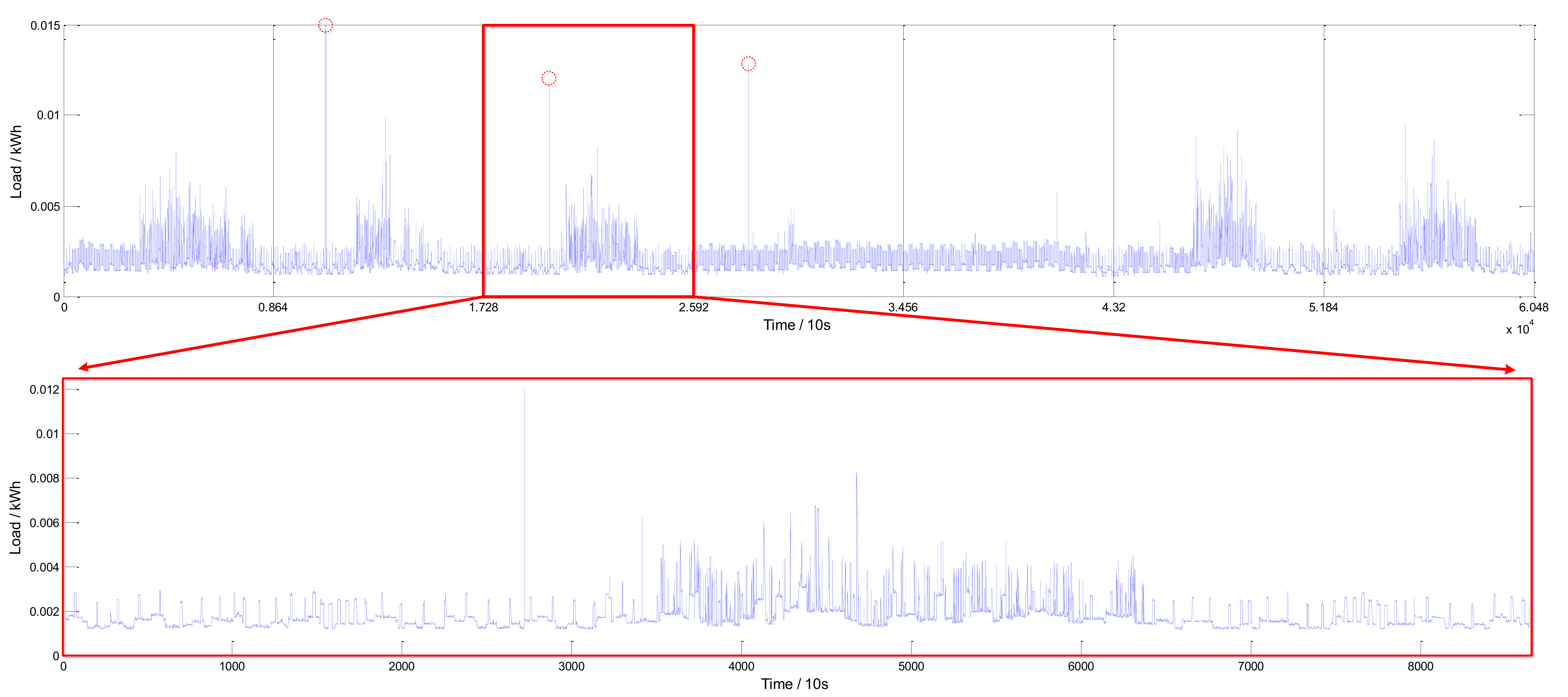}
\caption{One-week and one-day load data collected via the energy monitoring platform. \nop{\pei{the highlighted corrupted data points are hard to read in black-white printout.}}}\label{fig_realdataExp}
\end{center}
\end{figure}

\begin{table*}[t]	
\centering
	\caption{Results of corrupted data identification on real data: our appliance-driven approach vs. B-spline smoothing}
	\label{tab:resultComparison}	
		\begin{small}
		\begin{tabular}{|c|c|c|c|c||c|c|c|c|}
			\hline
			& \multicolumn{4}{c||}{\textbf{appliance-driven approach}}& \multicolumn{4}{c|}{\textbf{B-spline Smoothing}}\\
      \cline{2-9}
										&  $w = 1$ & $w = 2$ & $w = 3$ & $w = 5$ &  $df = 128$ & $df = 188$ & $df = 258$ & $df = 388$  \\
			\hline
			$Precision$ & $89.29\%$ & $95.83\%$ & $85.29\%$ & $84.38\%$ & $48.68\%$ & $50.00\%$ & $51.39\%$ & $47.44\%$  \\
			\hline
			$Recall$    & $50.00\%$ & $46.00\%$ & $58.00\%$ & $54.00\%$ & $72.55\%$ & $74.51\%$ & $82.35\%$ & $72.55\%$ \\
			\hline
			$F-measure$ & $64.10\%$ & $62.16\%$ & $69.05\%$ & $65.85\%$ & $58.27\%$ & $59.84\%$ & $60.16\%$ & $57.36\%$  \\			
			 \hline
		\end{tabular}
		\end{small}
\end{table*}

\begin{figure*}[t]
\begin{center}
\includegraphics[width=0.8\textwidth]{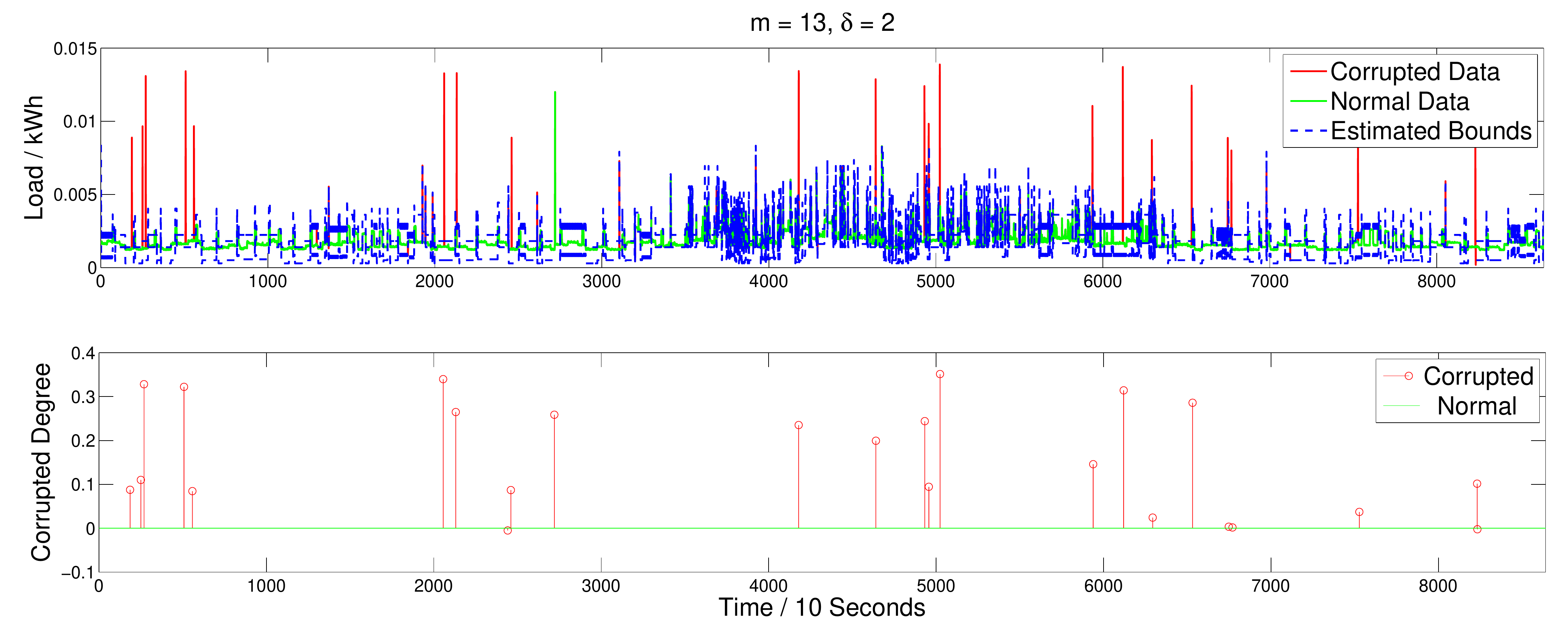}
\caption{Result of corrupted data identification on real data with our appliance-driven method ($w= 1, \delta = 2$); Estimated bounds denote the upper and lower power bounds based on current state vector; Corrupted degree indicates the value of the virtual appliance (Section~\ref{sec:systemModel}.)}\label{fig_outputOpt_m13d2}
\end{center}
\end{figure*}

In order to introduce more corrupted data, we asked three students to distort the load data with ``falsification'', \textit{i.e.}, they were asked to arbitrarily modify the aggregated load data within the range of $[0,\infty)$. These changed data together with the original corrupted ones were labeled and used as the ground-truth to verify the performance of our method.


Since the existing appliance-oblivious load data cleansing methods, such as B-spline smoothing, detect outliers and consider outliers as corrupted data, we use the terms ``outliers'' and ``corrupted data'' interchangeably hereafter. For outlier detection, four statistical results can be obtained: (1) true positive ($TP$), the number of points that are identified correctly as outliers; (2) false positive ($FP$), the number of points that are normal but are identified as outliers; (3) true negative ($TN$), the number of points that are normal and are not identified as outliers; (4) false negative ($FN$), the number of points that are outliers but are not identified. Using $TP, FP, TN$ and $FN$, we evaluate the following three broadly-used performance metrics: precision, recall, and F-measure. Precision is the ratio of the number of correctly detected corrupted values over the total number of detected values; recall is the ratio of the number of correctly detected values over the number of pre-labeled corrupted values; and the F-measure is a harmonic mean of precision and recall, \textit{i.e.},
\begin{equation}
\textit{F-measure}=\frac{2\cdot Precision\cdot Recall}{Precision+Recall}.
\end{equation}

For comparison, we implement and test an appliance-oblivious data cleansing method, B-spline smoothing, which is introduced in~\cite{chen2010automated} to identify corrupted load data. In the B-spline smoothing method, we set the confidence coefficient $\alpha=0.05$, which results in a confidence interval of $95\%$.  We treat the degree of freedom ($df$) as a variable, whose value is trained when smoothing the load curve data. For our method, the \emph{overlap index} is obtained as $O \approx 2$, and the upper bound of on-off switching events of appliances within the sampling interval is set to $2$, \textit{i.e.}, $\delta = 2$. According to Equation~(\ref{equ:windowSize}), the local window size, \textit{i.e.}, the value of $w$ in Algorithm~\ref{alg_seqentialOpt}, is set to $3$. Since the value of local window size is an estimation, in order to obtain more comprehensive performance evaluation for our method, we also vary the local window size in a range. 

Table~\ref{tab:resultComparison} summarizes some of the results from the two methods. Furthermore, Fig.~\ref{fig_outputOpt_m13d2} and Fig.~\ref{fig_outputBS258} illustrate one of the outcomes from our appliance-driven method and the B-spline smoothing method, respectively. From the results, we have the following interesting observations.

\begin{figure*}[t]
\begin{center}
\includegraphics[width=0.8\textwidth]{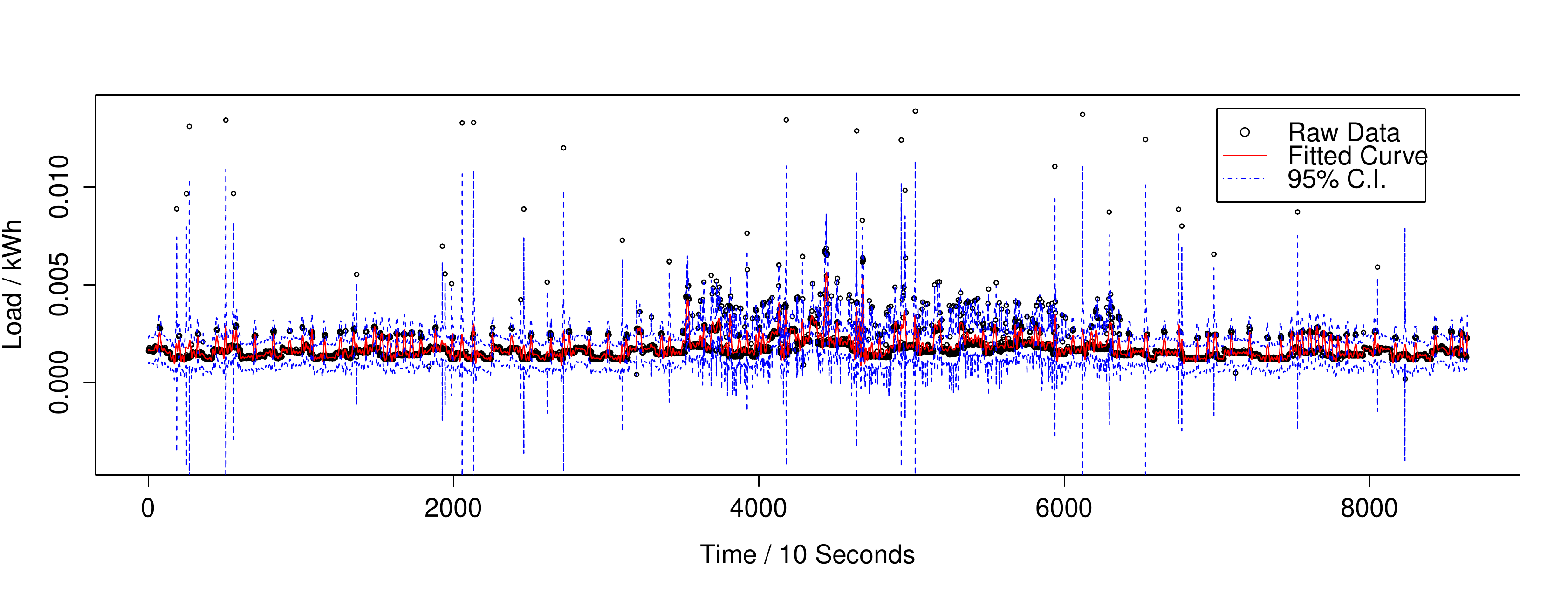}
\caption{Result of corrupted data identification on real data with the B-spline smoothing method ($df = 258$)}\label{fig_outputBS258}
\end{center}
\end{figure*}

\begin{itemize}
\item Comparing to B-spline smoothing, our method performs much better in \emph{Precision}, but worse in \emph{Recall}. This shows that our method can identify the corrupted data more accurately, even though our output does not cover the completed set of all corrupted data. In addition, our appliance-driven method achieves a higher \emph{F-measure}. \emph{F-measure} reflects a balanced mean between precision and recall, indicating that our method has overall better performance.   
\item The performance of our method remains roughly the same when the local window size is beyond the minimum value estimated using Lemma~\ref{lemma:1}. Further increase of the local window size does not bring clear performance gain but with a cost in longer running time. This suggests that our previous estimation on the minimum local window size for SLOA is appropriate. 
\end{itemize}

\section{Evaluation on Synthetic Data}\label{sec:evaluation2}

To thoroughly test our method, we evaluate its performance using large-scale synthetic data that simulates a large number of appliances and much diverse energy patterns. With different synthetic datasets, we can also test the robustness of our method.  

\subsection{Load Data Generation via Monte Carlo Simulation}
There is no standard model for the load curve data of a house, since the data actually results from a complex process related to human activities. We thus use the \emph{Monte Carlo} simulation to generate the load data using the following method:   
\begin{itemize}
	\item Given the lowest appliance power ($P_{min})$ and the highest appliance power ($P_{max}$), the lower bound of an appliance ($p_l$) is a random variable uniformly distributed between $P_{min}$ and $P_{max}$.  The upper bound of the appliance ($p_u$) is determined by a parameter called \textit{power range ratio} ($r$) and is calculated by $p_u = \min\{p_l + random([0, rp_l]), P_{max}\},$ where $random([0, rp_l])$ returns a random number uniformly distributed in the range $[0, rp_l]$. 
	\item At a given sampling frequency, each appliance reports its current power value, which is a random number uniformly distributed between the appliance's lower power bound and upper power bound. It reports $0$ if its state is \textit{off}.
	\item In a sampling interval, the number of total on-off switch events follows a Poisson distribution\footnote{Poisson distribution is a good model for situations where the total number of items is large and the probability that each individual item changes its state is small. It has been broadly adopted to simulate events related to human behavior, such as the number of telephone calls in a telephone system and the number of cars on high way.} with parameter $\lambda$.  
	\item At the end of each sampling interval, the load data of the house is recorded as the aggregated power value of all appliances (i.e., the sum of all appliances' load values).  
\end{itemize}

To introduce some corrupted data and test the effectiveness of our method, we ``corrupt" some data values by replacing them with random values uniformly distributed between $[0, Max]$, where $Max$ is a given large constant. The time interval of introducing corrupted data is assumed to follow an exponential distribution with the mean value of $\mu$. 

\subsection{Corrupted Data Identification on Large-Scale Appliances}

The parameters used to generate the synthetic data and the corrupted data are listed in Table~\ref{tab:parameter}. 
\begin{table}[t]
	\caption{Parameter settings for load data generation and corruption}\label{tab:parameter}
	\centering
	\begin{small}
		\begin{tabular}{|c|c|}
			\hline
			\emph{Parameter} & \emph{Setting}\\
			\hline
			Number of Appliances ($m$) & $50$ \\
			\hline
			Sampling Frequency($f$) & $1/6 Hz$ \\
			\hline
			Total Time Span & $3600 s$ \\
			\hline
			Lowest Appliance Power($P_{min}$) & $50 w$ \\
			\hline
			Highest Appliance Power($P_{max}$) & $2000 w$ \\
			\hline
			Power Range Ratio($r$) & $15\%$ \\
			\hline
			Initial State($S$) & $[0,0,\cdots,0]^{T}$ \\
			\hline
			Poisson Parameter($\lambda$) & $5$ \\
			\hline
			Exponential Parameter($\mu$) & $30$ \\
			\hline
			Corrupted Data Range & $[0, 50kW]$ \\
			\hline
		\end{tabular}
	\label{tab_SystemParameterSetting}
	\end{small}
\end{table}

\begin{table*}[t]
	\caption{Results of corrupted data identification on synthetic data: appliance-driven method vs. B-spline smoothing method}\label{tab:resultComparison2}
	\centering
	\begin{small}
		\begin{tabular}{|c|c|c|c||c|c|c|c|}
			\hline
			& \multicolumn{3}{c||}{\textbf{appliance-driven Method}}& \multicolumn{4}{c|}{\textbf{B-spline Smoothing}}\\
      \cline{2-8}
			 &  $\delta = 4$ & $\delta = 5$ & $\delta = 6$ &  $df = 140$ & $df = 160$ & $df = 180$ & $df = 200$  \\
			\hline
			$Precision$  & $93.94\%$ & $93.94\%$ & $100\%$ & $78.57\%$ & $86.49\%$ & $84.61\%$ & $84.21\%$  \\
			\hline
			$Recall$  & $81.58\%$ & $81.58\%$ & $63.16\%$ & $86.84\%$ & $84.21\%$ & $86.84\%$ & $84.21\%$  \\
			\hline
			$F-measure$   & $87.32\%$ & $87.32\%$ & $77.42\%$ &  $82.50\%$ & $85.33\%$ & $85.71\%$ & $84.21\%$  \\
			\hline
		\end{tabular}
		\end{small}
\end{table*}

We treat the bound on the total number of on-off switches in a sampling interval $\delta$ as a variable. To speed up the processing, we set the local window size to $1$. The small local window size may not lead to the best performance of SLOA.  However, as shown in our experimental results, even with this setting, our method already performs better than B-spline smoothing. For the B-spline smoothing method, the degree of freedom ($df$) is set as a variable and is trained when smoothing the synthetic data.

The performance results of our method and the B-spline smoothing method are summarized in Table~\ref{tab:resultComparison2}. Fig.~\ref{fig_outputOpt_m50d5} and Fig.~\ref{fig_outputBS160_sim} illustrate one of the outcomes from our method and the B-spline smoothing method, respectively.

From the results, we can see that the our method works effectively on large-scale synthetic data. In particular, we find that the \textit{precision} of our method increases with increase of $\delta$, and can even reach $100\%$. This result indicates that our method can provide excellent correct identification when $\delta$ is large enough. Regarding the overall performance in view of \textit{F-measure}, our method works better with a smaller $\delta$ value and outperforms B-spline smoothing clearly. 

\nop{\pei{Do you think we need (or can) do an ROC analysis?}}

\begin{figure}[t]
\begin{center}
\includegraphics[width=3.4in,height=1.8in]{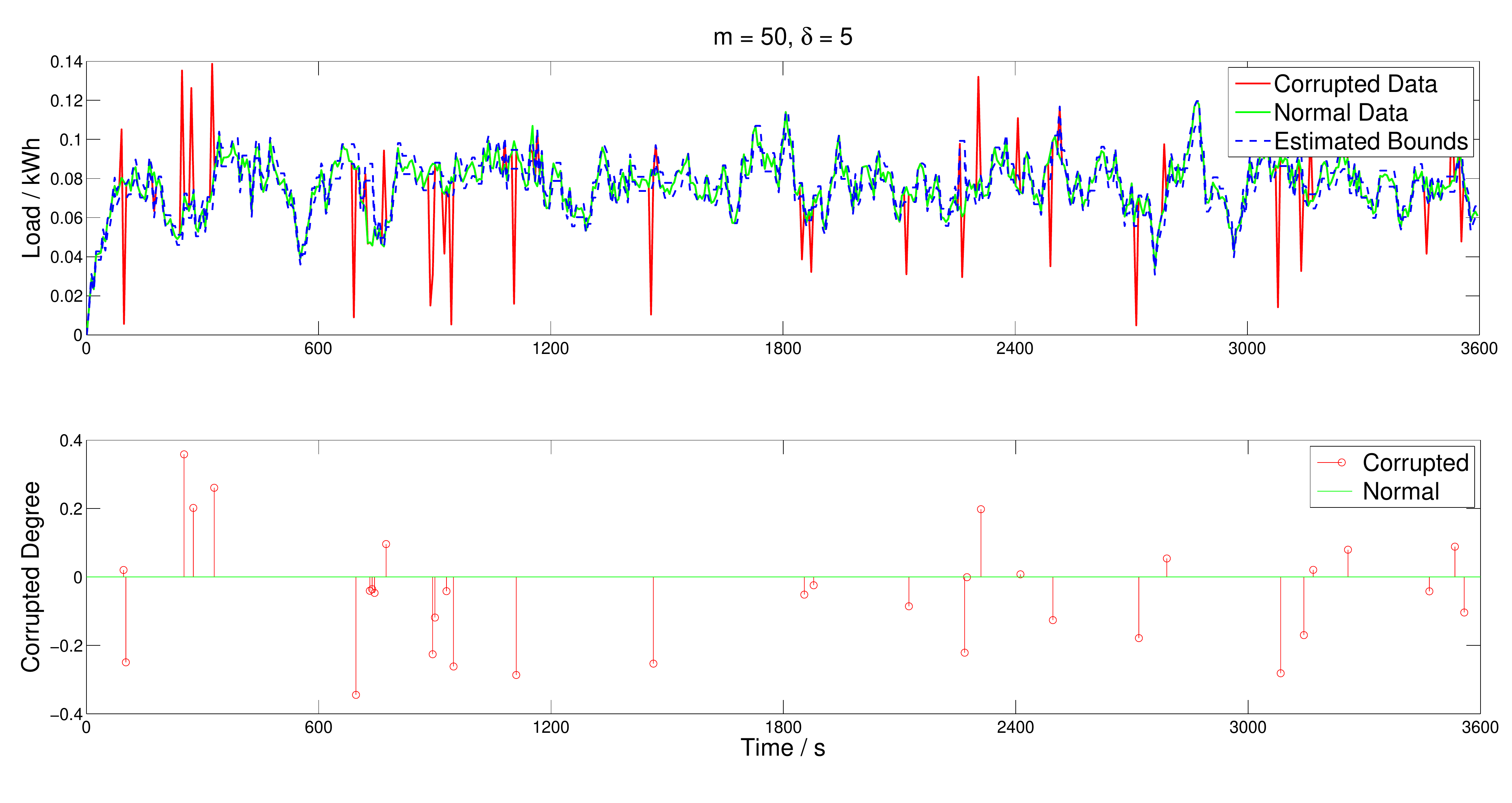}
\caption{Result of corrupted data identification on synthetic data with our appliance-driven method ($w = 1, \delta =5$)}\label{fig_outputOpt_m50d5}
\end{center}
\end{figure}
\begin{figure}[t]
\begin{center}
\includegraphics[width=3.4in,height=1.3in]{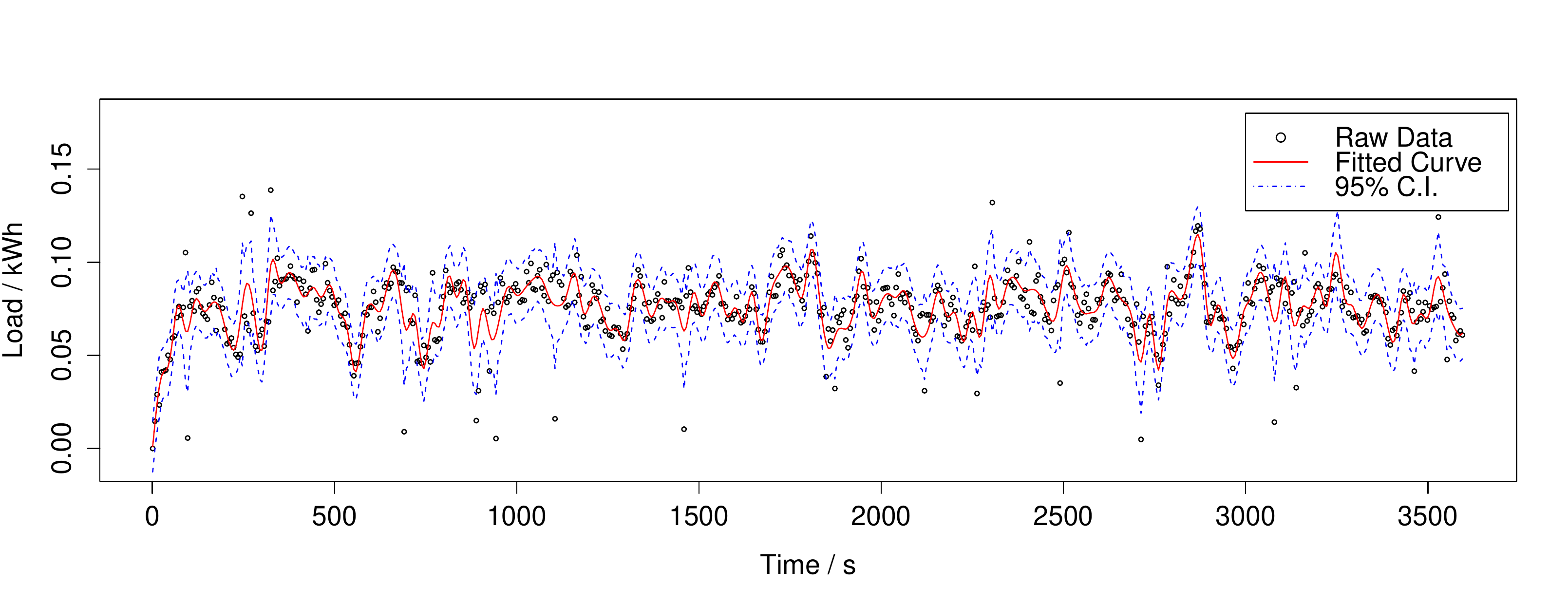}
\caption{Result of corrupted data identification on synthetic data with B-spline smoothing method($df = 160$)}\label{fig_outputBS160_sim}
\end{center}
\end{figure}

\subsection{Identification of Consecutive Corrupted Data}
In practice, we often meet the situation that all data within a small time interval are corrupted or lost. Consecutive corrupted data poses a big challenge to regression-based methods, as will be illustrated in this subsection. 

To introduce consecutive corrupted data, we replace the load data in a small time window as $0$s, as shown in the upper part of Fig.~\ref{fig:miss1}. We then use our method and the B-spline smoothing method to test the data. Fig.~\ref{fig:miss1} and Fig.~\ref{fig:miss2} illustrate one outcome from our appliance-driven approach and the B-spline smoothing method, respectively.  

From the results, we can see that our method does much better than B-spline smoothing for consecutive corrupted data identification. With $\delta = 5$, our method can correctly identify all the corrupted data. On the other hand, even though we regulate the parameters for B-spline smoothing, it almost failed every time to identify even half of the corrupted data. 

\begin{figure}[t]
\begin{center}
\includegraphics[width=3.4in,height=1.8in]{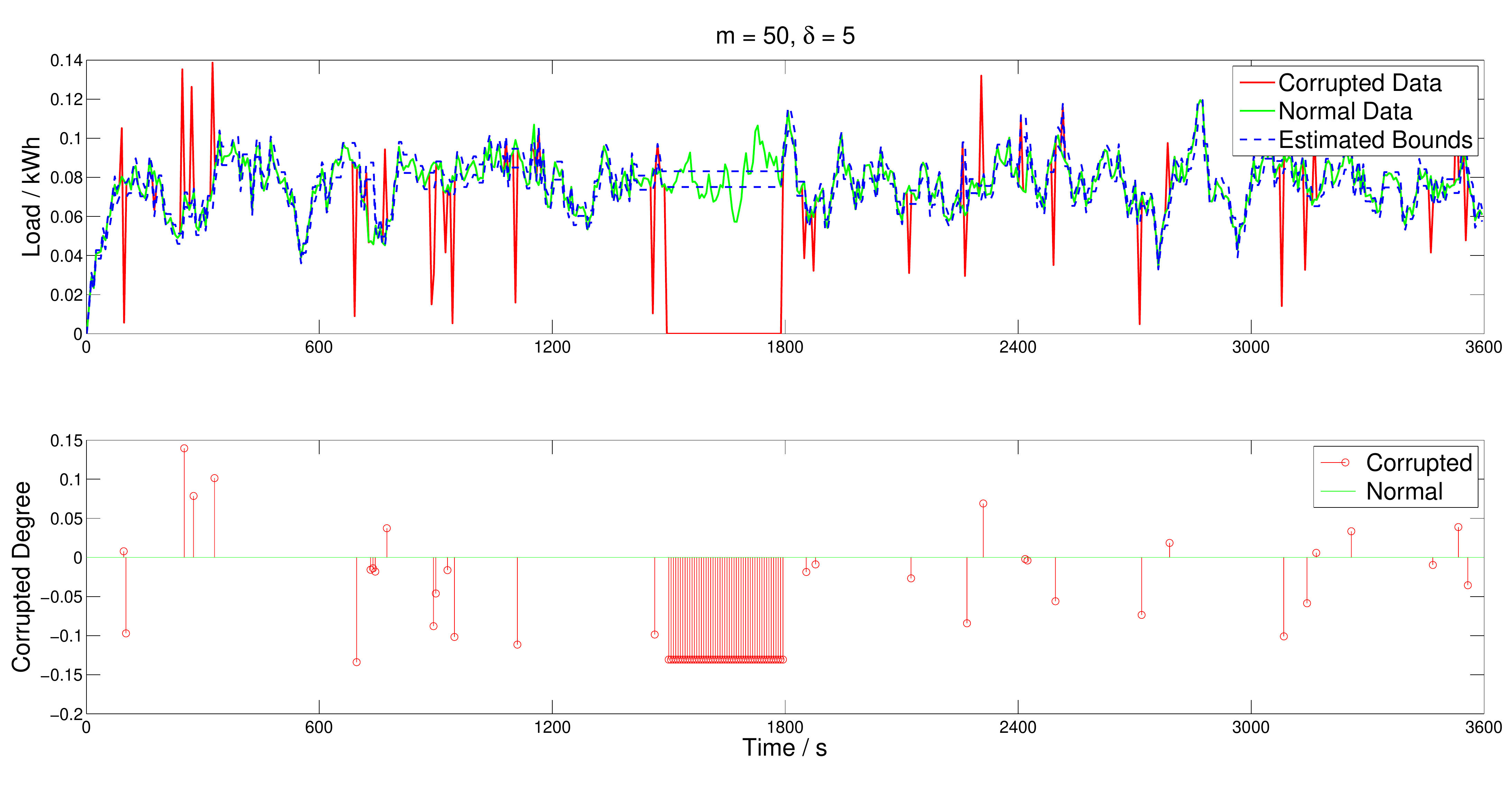}
\caption{Identification of consecutive corrupted data with our appliance-driven method}\label{fig:miss1}
\end{center}
\end{figure}
\begin{figure}[t]
\begin{center}
\includegraphics[width=3.4in,height=1.3in]{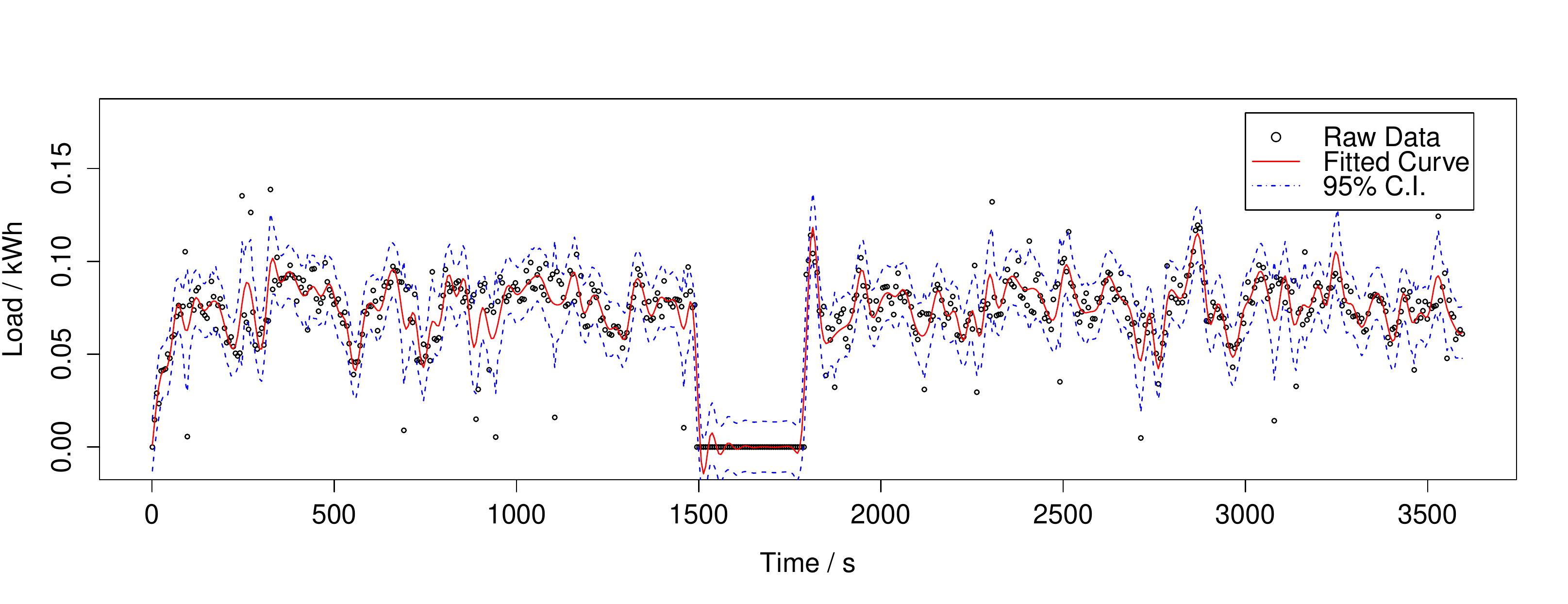}
\caption{Identification of consecutive corrupted data with B-spline smoothing method ($df = 100$)}\label{fig:miss2}
\end{center}
\end{figure}

An interesting phenomenon can be found around the consecutive corrupted data in Fig.~\ref{fig:miss2}. There is an apparent trend with B-spline smoothing to fit the corrupted data. This is mainly because the B-spline smoothing method tries to fit the curve pattern and reduce the total bias error with global optimization, indicating that it cannot deal with consecutive corrupted data well.   

\begin{table*}[t]
	\caption{Robustness tests with incorrect power ranges of appliances}\label{tab:resultComparison3}
	\centering
	\begin{small}
		\begin{tabular}{|c|c|c|c|c||c|c|c|c|}
			\hline
			& \multicolumn{2}{c|}{\textbf{Widen ($5\%$)}}& \multicolumn{2}{c||}{\textbf{Widen ($10\%$)}} & \multicolumn{2}{c|}{\textbf{Shift\& Widen ($5\%$)}} & \multicolumn{2}{c|}{\textbf{Shift\& Widen ($10\%$)}}\\
      \cline{2-9}
			     & $\delta = 3$ & $\delta = 4$ & $\delta = 3$ & $\delta = 4$ & $\delta = 3$ & $\delta = 4$ & $\delta = 3$ & $\delta = 4$  \\
			\hline
			$Precision$  & $93.55\%$ & $87.50\%$ & $91.30\%$ & $94.12\%$ & $67.39\%$ & $90.91\%$ & $71.11\%$ & $84.62\%$ \\
			\hline
			$Recall$  & $76.32\%$ & $55.26\%$ & $55.26\%$ & $42.11\%$ & $81.58\%$ & $78.95\%$ & $84.21\%$ & $57.89\%$ \\
			\hline
			$F-measure$  & $84.06\%$ & $67.74\%$ & $68.84\%$ & $58.18\%$ & $73.81\%$ & $84.51\%$ & $77.11\%$ & $68.75\%$ \\
			\hline
		\end{tabular}
		\end{small}
\end{table*}

\section{Robustness Testing} \label{sec:robust}

One may question whether the performance of SLOA relies on a correct initial state vector, accurate information regarding appliances power ranges, and an accurate estimation on appliances' on-off states. all of such information may be hard to obtain in practice. To answer this question, we test the robustness of SLOA. We use the synthetic data created using the same parameters in Table~\ref{tab:parameter}. We first disclose the test results and then explain the reasons. 

\subsection{Impact of the Initial State}

For this test, we change the initial state of an appliance to a random $0$-$1$ value, and perform multiple tests. Fig.~\ref{fig_initialState} shows one of the outcomes. We find that, even with an incorrect initial state, our method can always recover to correct load data after a few steps. This result indicates that our SLOA method is robust against inaccurate initial power state setting. 

\begin{figure}[t]
\begin{center}
\includegraphics[width=3.2in,height=1.2in]{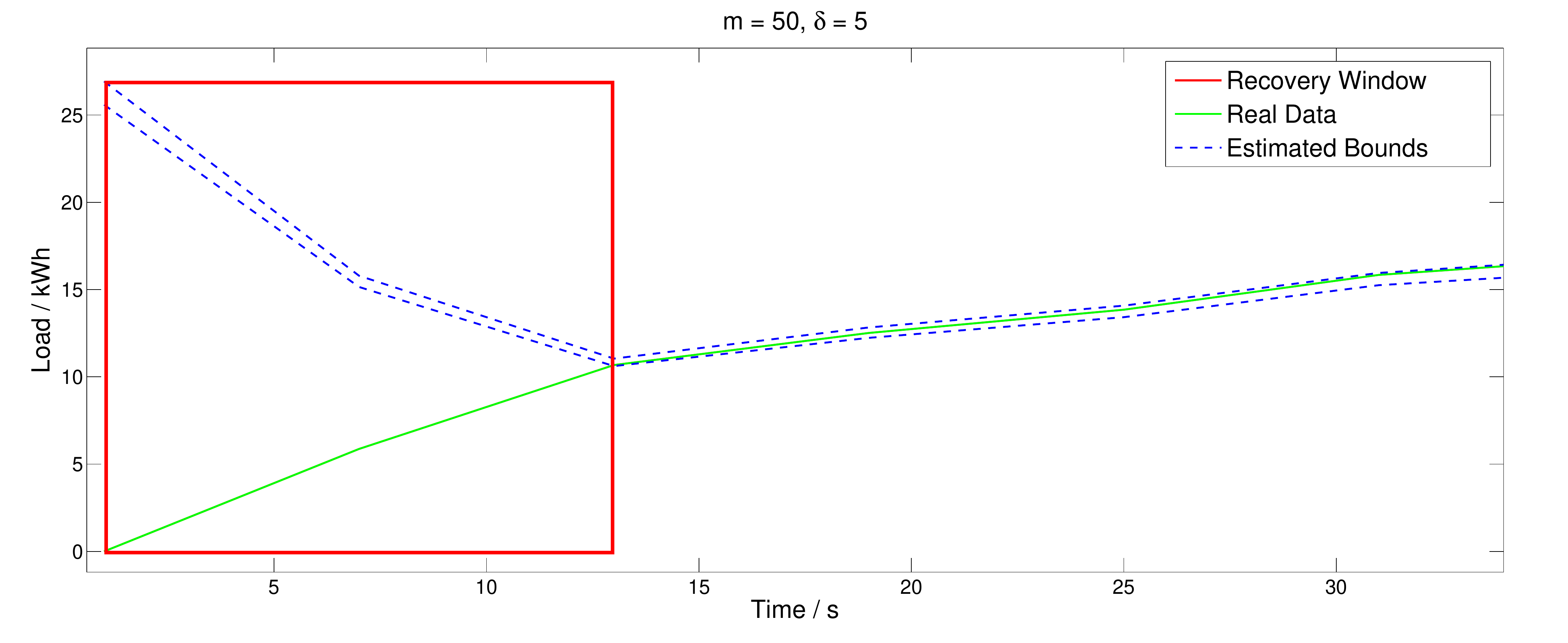}
\caption{Fast recovery of estimated load starting from a random initial state}\label{fig_initialState}
\end{center}
\end{figure}

\subsection{Impact of Power Ranges}

In practice, we may not precisely know the power ranges of appliances. Based on this consideration, we run extra simulations to test the robustness of our method when the power range information of appliances is inaccurate. We carry out two kinds of tests as follow.
\begin{itemize}
\item Widen power range: each appliance's power range is widened by $5\%$ or $10\%$, respectively, with the center power value, \textit{i.e.}, $(\text{upper bound} - \text{lower bound})/2$, unchanged.
\item Shift\& widen power range: each appliance's lower power bound is increased by $5\%$ or $10\%$, and upper bound is increased by $5\%$ or $10\%$, respectively. Note that the above operations will shift the center power value as well as widen the power range. 
\end{itemize}
We do not consider the situation where the appliances' power ranges are narrowed, since intuitively a user can always widen an appliance's power range if she/he is not sure about the right values. 

The test results are summarized in Table~\ref{tab:resultComparison3}. The results clearly indicate that, with inaccurate or even wrong power ranges of appliances, our method can still manage to identify corrupted data with high precision. 

\subsection{Impact of State Vector}
We have seen that our method can give correct bounds for energy consumption most of the time. Accordingly, we might infer that the estimated states of the appliances should be the same with the real situation, or at least quite close.

In order to verify this conjecture, we calculate the difference (one-norm distance) between the estimated state $S_{e}$ and the corresponding real state $S_{r}$ at each time instance. Fig.~\ref{fig_innerState} shows the result. 

\begin{figure}[t]
\begin{center}
\includegraphics[width=3.2in,height=1.2in]{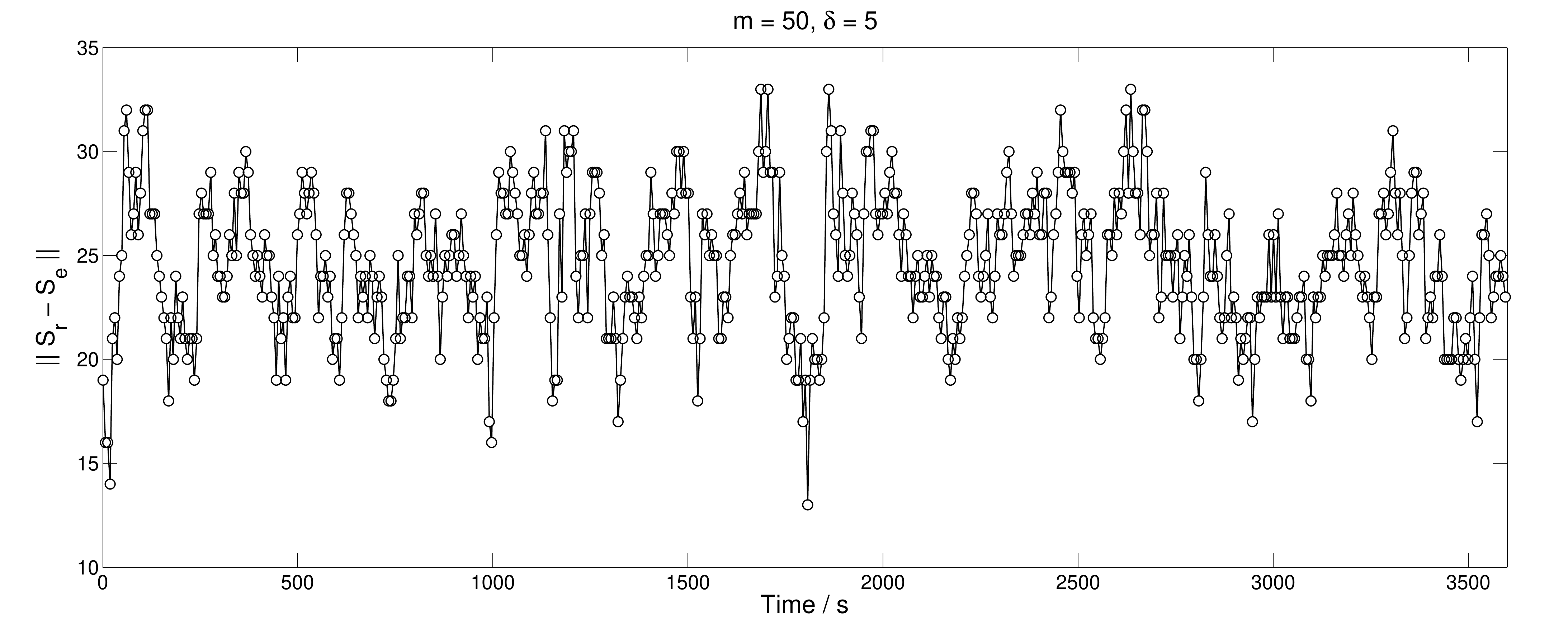}
\caption{Difference between estimated state $S_{e}$ and real state $S_{r}$}\label{fig_innerState}
\end{center}
\end{figure}

To our surprise, the estimated states are not close to the real states, and actually deviate remarkably from their real states. We can see that in Fig.~\ref{fig_innerState}, the mean distance between $S_{e}$ and $S_{r}$ is around $25$, indicating that nearly half of the appliances are not estimated with the correct states. This shows that the solution to the CDIP problem is not unique but multiple, and our method can provide the right load data without need to always find the right states of appliances. 

\subsection{Why Is SLOA Robust?}

In real life, a lot of appliances are with similar or overlapped power range. In this sense, we indeed do not need to know the exact state for similar appliances as long as we can give a good approximation for their total consumption. In addition, due to the temporal sparsity of on-off switch events in the short sampling interval and the fact that only some appliances are on at any time instant, the negative impact of inaccurate power range estimation on one appliance can be offset by the negative impact of incorrect state estimation of another appliance. The offsetting is enforced automatically by the optimization objective function that minimizes the gap between the actual load data and the estimated value.             


\section{Conclusion and Future Work}\label{sec:conclusion}

To answer the industrial call for improving quality of load data, we developed a new appliance-driven approach for corrupted data identification that particularly takes advantage of information available on the demand side. Our appliance-driven approach considers the operating ranges of appliances that are readily available from users' manual, technical specification, or public websites~\cite{EnergyStar2013}. It identifies corrupted data by solving a carefully-designed optimization problem. To solve the problem efficiently, we developed a sequential local optimization algorithm (SLOA) that practically approach the original NP-complete problem approximately by solving an optimization problem in polynomial time.      

We evaluated our method using both real trace data from a real-world energy monitoring system and large-scale synthetic data. Test results indicate that our method can precisely capture corrupted data. In addition, SLOA is robust under various test scenarios, and its performance is resilient to inaccurate power range information or inaccurate power state estimation. 

Our method greatly augments the arsenal of existing load data cleansing tools to minimize human effort in identifying corrupted data. Yet, we ignore the privacy issues in this study. Even if the customers and the utility companies have an aligned common goal for accurate load data, some customers may be reluctant to collaborate due to privacy concerns. Our future research is to enhance our appliance-driven approach by developing privacy-preserving load data cleansing methods. In addition, how to replacing aberrant values and missing values is out of the focus of this paper, because this issue is relevant to utilities' internal rules and thus requires human interaction. Considering various policies and methods for load data imputation will be our future work. 

\nop{
\section*{Acknowledgment}
Some words on funding support ...
}
\balance
\small \baselineskip 10pt
\bibliographystyle{abbrv}
\bibliography{Reference}




\section*{Appendix A: Proof of NP-Completeness of CDIP}\label{sec:appendixA}

\subsection{Preparation}
First, we introduce a tree structure $T = (M, N)$, which is a \emph{complete} $M$-ary tree with \emph{height} of $N$, \textit{i.e.}, every internal node has exactly $M$ children and all leaves have the same depth of $N$. Furthermore, each edge $(i,j)$ of $T$ has a non-negative cost $c(i,j)$, which will be defined later.  

Assume that load data $\{y_1, y_2, \cdots, y_n\}$ is generated by $m$ appliances with the initial state vector $S_0$. Also assume that the upper bound on the total number of on-off switches within a sampling interval is $\delta (<m)$. We can build the following tree:

\begin{enumerate}
\renewcommand{\theenumi}{Step \arabic{enumi}}
 \item Set $S_0$ as the root of the tree.
 \item Set the children of root as all possible states that can be transited from $S_0$, with the constraint that the total number of on-off switches is no larger than $\delta$.  Therefore, we can add $M$ children to the root, where $M =C^{0}_{m}+C^{1}_{m}+\cdots+C^{\delta}_{m}$;
 \item\label{repeat} For each node of the tree, set its children as all possible states ($M$ states) that can be transited from it;
 \item Repeat~\ref{repeat}) from $t=1$ to $n$. At the end, we obtain $T = (M, N)$, where $N = n$; 
 \item Set the cost of edge $(i, j)$ as $c(i, j) =\left| v \right|$, where $v$ is obtained by solving the optimization problem:\\  
minimize $\left| v \right|$, subject to $\left(P^{T}_{l}S_{j} - v\right){/f}\leq y_i\leq \left(P^{T}_{u}S_{j} + v\right){/f}$.
\end{enumerate}

Thus, we can translate CDIP into the problem of finding the minimum-cost path in $T(M, N)$ from the root to a leaf. Equivalently, we need to answer the following question: given a constant $k$, is there a path in $T(M,N)$ from the root to a leaf with total cost no larger than $k$? In the following, we call a path from the root to a leaf in the tree as a \textit{full path}. 

With the notation above, CDIP can be re-formulated as
\begin{equation*}
\begin{split}
 CDIP = \{ \left\langle T, c, k \right\rangle:
& T = (M, N),\\
& c \text{ is the cost function },\\
& k \in \Re^{+}, \text{ and }\\
& T \text{ has a full path with cost } \leq k\}.
\end{split}
\end{equation*}

We next reduce a well-known NP-complete problem, the \emph{Traveling Salesperson Problem }(TSP) to CDIP. TSP can be formulated as
\begin{equation*}
\begin{split}
 TSP = \{ \left\langle G, c', k \right\rangle:
& G = (V, E) \text{ is a complete graph },\\
& c' \text{ is the cost function },\\
& k \in \Re^{+}, \text{ and }\\
& G \text{ has a Hamiltonian cycle with cost } \leq k\}.
\end{split}
\end{equation*}

\subsection{Proof}
%

We complete the proof in two steps: firstly we show that CDIP is NP; then, we prove that CDIP is NP-hard by showing $TSP \leq_{P} CDIP$, \textit{i.e.}, there exists a reduction from TSP to CDIP.

\emph{a. CDIP is NP}
\begin{itemize}
\item \emph{Certificate:} A path of $T$.

\item \emph{Algorithm:}
\begin{itemize}
\item Check that the path is full, \textit{i.e.}, the path starts from the root and ends at a leaf.
\item Sum up the edge costs along the path and check if it is no larger than $k$.
\end{itemize}

\item \emph{Polynomial Time:}
We need $N$ steps to check the fullness of path and obtain the total cost.
\end{itemize}

\emph{b. CDIP is NP-hard}
\begin{itemize}
\item Firstly, we develop an algorithm $F:\left\langle G, c', k \right\rangle \rightarrow \left\langle T, c, k \right\rangle$, \textit{i.e.}, $G$ and $c'$ in TSP can be transferred to $T$ and $c$ in CDIP as follow:

\begin{enumerate}
\renewcommand{\theenumi}{Step \arabic{enumi}}
\item Choose any node of $G$ as the root of $T$;
\item\label{repeat2} For each leaf node of the current tree, add its children as all the other nodes of $G$. Since $G$ is a complete graph, we can add $\left|V\right|-1$ children to each leaf node, where $\left|V\right|$ is the number of nodes in $G$.
\item Repeat~\ref{repeat2} for $\left|V\right|$ times. At the end, we build $T = (M, N)$, where $M = \left|V\right|-1$ and $N =\left|V\right|$; 
\item\label{special} Set the cost of edge $(i, j)$ in $T$, $c(i,j)$, as follows: 
\begin{enumerate}
\item Initialization: $c(i,j) = c'(i,j)$, where $c'(i, j)$ is the edge cost in $G$.
\item For each edge $(i,j)$ of $T$ where $j$ is a non-leaf node, if $j$ has appeared in the path from the root (including the root) to $i$, \textit{i.e.}, $j$ is an ancestor of $i$ in the tree already, replace $c(i, j)=\infty$.   
\item For each edge $(i,j)$ of $T$ where $j$ is a leaf node, if $j$ is not the same as the root node, replace $c(i, j)=\infty$.
\end{enumerate} 
\end{enumerate}

To help understand the construction of $T$ with $G$, Fig.~\ref{fig:example} show an example with three nodes in $G$. 

\begin{figure}[t]
\begin{center}
\includegraphics[width=3in,height=1in]{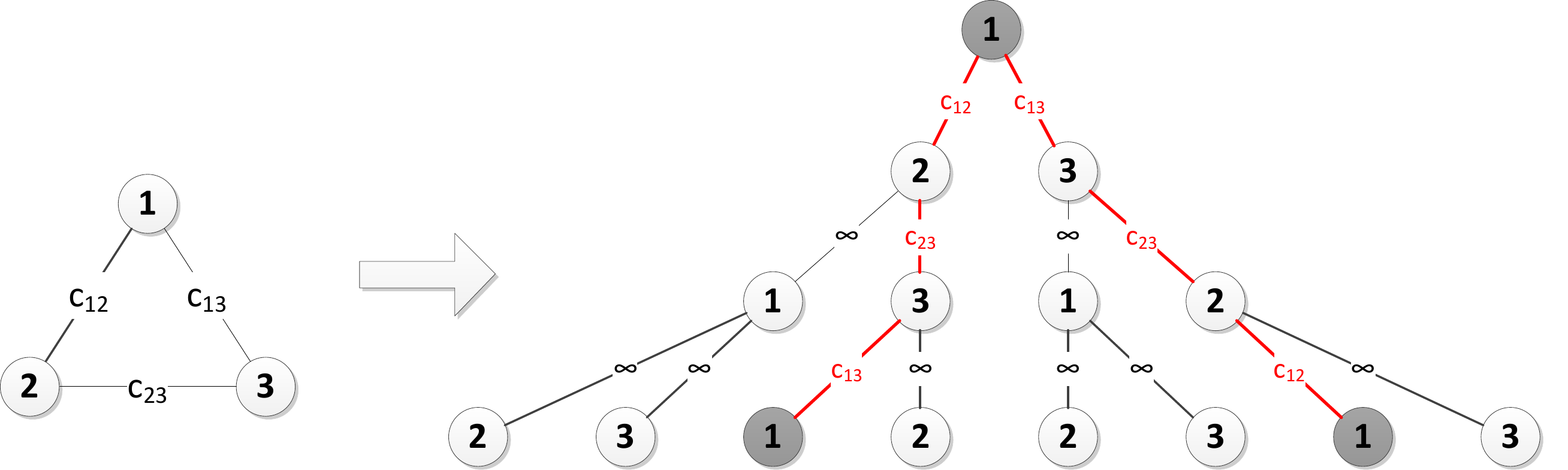}
\caption{An example showing the construction of $T$ with $G$}\label{fig:example}
\end{center}
\end{figure}

\item Secondly, it is easy to see that $F$ takes $O(N^2)$ running time. 

\item Thirdly, we show that 

\begin{equation*}
\left\langle G, c', k \right\rangle \in TSP \Leftrightarrow \left\langle T, c, k \right\rangle \in CDIP.
\end{equation*}

\begin{itemize}
\item $(\Rightarrow)$
\begin{equation*}
\begin{split}
    &G \text{ has a Hamiltonian cycle with cost} \leq k.\\
		&\Rightarrow \text{there exists a full path in tree} T \text{ with cost } \leq k. \\ 
		&\text{(Note that there will be no internal node along the path} \\ 
		&\text{occurring more than once, otherwise the cost will}\\
		&\text{be infinite with operation in~\ref{special}.) }\\
\end{split}
\end{equation*}
\item $(\Leftarrow)$
\begin{equation*}
\begin{split}
    &T \text{ has a full path with cost } \leq k.\\
		&\Rightarrow \text{there exists a traverse instance in its corresponding}\\
		&\text{graph } G \text{ with cost } \leq k.\\
		&\text{(Note that based on the tree construction procedure, }\\
		&\text{only the full paths starting and ending at the same}\\
		& \text{node can have a cost no larger than } k, \text{because other} \\
		&\text{paths have a cost of infinity.)}\\
		&\Rightarrow \text{so } G \text{ has a Hamiltonian cycle with cost } \leq k.
\end{split}
\end{equation*}
\end{itemize}
\end{itemize}
With \emph{step a. }and \emph{step b.}, we prove that CDIP is NP-complete.

\section*{Appendix B: Computational Complexity}\label{sec:appendixB}

For problem (Equation~\ref{minimization3}), the second constraint means that during a sampling interval, there are at most $\delta$ out of $m$ appliances that can change their states. This results in ${{m}\choose{0}}+{{m}\choose{1}}+\cdots+{{m}\choose{\delta}}$ feasible solutions. Therefore, the total number of possible state sequences is $M^{n}$, where $M={{m}\choose{0}}+{{m}\choose{1}}+\cdots+{{m}\choose{\delta}}$. Thus, the computational complexity of problem (Equation~\ref{minimization3}) is $O(M^{n})$, which is exponential.

As to problem (Equation~\ref{minimization4}), the whole search space is split into $n$ local windows of size $w$, and the optimization is confined within the local window. Given $S_{i-1}$, there are at most ${{m}\choose{0}}+{{m}\choose{1}}+\cdots+{{m}\choose{\delta}}$ instances. We have to traverse all the instances to find the one that minimizes $v_i$ in each step of a local optimization. Therefore, the computational complexity to find a local optimal solution with $w$ steps is $O(M^{w})$, where $M={{m}\choose{0}}+{{m}\choose{1}}+\cdots+{{m}\choose{\delta}}$.

Hence, after applying SLOA in each of the $n$ local windows, the total computational complexity to obtain the final solution is $O(n \cdot M^{w})$. Considering that the number of appliance $m$ is a constant value and $w$ is also a small constant, SLOA cuts down the computational complexity of the original problem from exponential to polynomial.

\end{sloppy}
\end{document}